\documentstyle[graphicx,longtable]{mn2e}

\newcommand{\hMpc}{{\ifmmode{h^{-1}{\rm Mpc}}\else{$h^{-1}$Mpc}\fi}}
\newcommand{\hkpc}{{\ifmmode{h^{-1}{\rm kpc}}\else{$h^{-1}$kpc}\fi}}

\def\approxlt{\mathrel{\spose{\lower 3pt\hbox{$\sim$}}
	\raise 2.0pt\hbox{$<$}}}
\def\approxgt{\mathrel{\spose{\lower 3pt\hbox{$\sim$}}
	\raise 2.0pt\hbox{$>$}}}
\def\approxpropto{\mathrel{\spose{\lower 3pt\hbox{$\sim$}}
	\raise 2.0pt\hbox{$\propto$}}}

\title{New Periodic Variables from the Hipparcos Epoch Photometry}
\author[C.Koen \& L.Eyer]{Chris Koen$^{1}$ \& Laurent Eyer$^{2, 3, 4}$\\
$^{1}$ South African Astronomical Observatory P O Box 9, Observatory,
     7935 Cape, South Africa\\
$^{2}$ Instituut voor Sterrenkunde, Katholieke Universiteit Leuven, Belgium\\
$^{3}$ Observatoire de Gen\`eve, 1290 Sauverny, Switzerland\\
$^{4}$ Princeton University Observatory, Princeton, NJ 08544, USA}
\date{Accepted --.
      Received -- ;
      in original form --}

\pagerange{\pageref{firstpage}--\pageref{lastpage}}
\pubyear{2001}

\begin{document}

\maketitle

\label{firstpage}

\begin{abstract}
Two selection statistics are used to extract new candidate periodic
variables from the epoch photometry of the Hipparcos catalogue. The primary
selection criterion is a signal to noise ratio. The dependence of this
statistic on the number of observations is calibrated using about 30 000
randomly permuted Hipparcos datasets. A significance level of 0.1\% is
used to extract a first batch of candidate variables. The second criterion
requires that the optimal frequency be unaffected if the data are
de-trended by low order polynomials. We find 2675 new candidate periodic
variables, of which the majority (2082) are from the Hipparcos ``unsolved"
variables. Potential problems with the interpretation of the data (e.g. 
aliasing) are discussed. 
\end{abstract}


\section*{\large \bf 1 INTRODUCTION}

The Hipparcos catalogues of variable stars (volumes 11 and 12 of ESA 1997)
arose from the work of two groups, one at the Geneva Observatory and the
other at the Royal Greenwich Observatory. Van Leeuwen (1997a) provided a
brief summary of the variable star identification methods used by the two
groups, while more detailed descriptions of the two independent strategies
are available in Eyer (1998) and van Leeuwen et al. (1997).  Although the
variable star data have been widely used in studies of individual cases,
or small collections of specific variables, little follow-up work with a
global approach to the Hipparcos epoch photometry has been attempted since
the release of the catalogues in 1997. One such study is the one by Koen
(2001) who searched through the Hipparcos database for stars having
multi-periodic behaviour.

Because of unavoidable statistical fluctuations, stars flagged as constant
could in fact be variable, or even periodic, near the level of the
precision of the Hipparcos measurements. The converse is inevitably also
true, stars flagged as variable could be in reality constant stars at the
Hipparcos precision. It is therefore important to have a sound evaluation
of these contaminations.

The methodologies of the original analyses 
relied in the first instance on the estimated standard errors of the
individual magnitudes. These errors were used for example in the selection
of variable stars, the determination of the limiting threshold to perform
a period search, and the determination of the intrinsic amplitudes of the
so-called unsolved variables 
(although details of the approaches taken by the two groups of
analysts differed -- see e.g. van Leeuwen 1997b).
Indeed, not all stars flagged as variable were searched for a periodic
behaviour -- only those satisfying a criterion depending on the estimated
amplitude, noise level and number of measurements were pursued. Such a
criterion was set in order to minimize the number of variables with
incorrect frequency determinations (due to aliasing). As noted by Eyer \&
Genton (1999), it is suspected that slight systematic shifts are present
in the estimated standard errors, especially for the bright and faint ends
of the catalogue.

The aim of the present study is to overcome the difficulties associated
with the estimation of the variable star detection errors, first by
tackling the problem from the outset by using Fourier methods, and second
by estimating the type I errors without making assumptions about the
nature of the data. The latter task is accomplished by determining the
general statistical behaviour of the signal to noise ratio (section 2),
which allows the computation of the expected number of spurious variables
in a sample of candidate periodic stars. It therefore allows us to set
numerical values on the Type I errors in order to keep these to an
acceptable level.

We computed power spectra for, and fitted sinusoids to, the observations
of 30349 stars in order to calibrate our primary variable selection
statistic, which is a signal to noise ratio.  The test was then applied
to the time series of 94336 stars out of a total of 118204 stars contained
in the Hipparcos database. (Stars already flagged as periodic, and
unflagged stars fainter than $V=10$ were excluded from consideration).

The question of estimation of errors is not the only reason why the study
we are presenting is of potential interest. Our method has the added
advantages of considering the Hipparcos data from an entirely different
viewpoint, and, of course, allows further extraction of relevant
information from the photometric data.

The basic philosophies of the original studies and the present one are
rather different. In the former, elimination of individual objects and an
iterative approach were used: visual inspection of light curves and all
phase diagrams, re-calculation of periodic solutions on an individual
basis and literature searches were done. The main aim of the original
study was to produce statistically very well confirmed periodic variable
stars. In this paper we develop simple but stringent criteria, and we
publish the output as it stands, without eliminating any results, however
unpalatable.
Furthermore, it is clear that the scrutiny of individual stars which
was carried out by the {\it Hipparcos} consortium (e.g. van
Leeuwen 1997a) is not viable for large scale photometric surveys
currently underway, or about to be begin. Fully automated algorithms,
such as those proposed here, are needed to deal with star counts 
which may be several orders of magnitudes greater than in the case of
{\it Hipparcos}.
 
The selection criteria are discussed in section 2. An alternative approach
to the setting of significance levels is indicated in section 3. Results
are given in section 4, and conclusions presented in section 5.

The interested reader is referred to ESA (1997), Eyer \& Grenon (2000)
and van Leeuwen (1997b), for discussions of quality and quality control of
the {\it Hipparcos} epoch photometric data.

\section*{\large \bf 2 THE SELECTION CRITERIA}

The primary variable selection criterion is based on the premise that the 
best-fitting
sinusoid for a true variable ought to have a higher amplitude than 
would be the case for a constant star with a similar noise level and number
of observations. Ideally, the test should be performed by, in a sense, 
comparing the data for a given star with itself, by using a permutation test. 
The latter is
performed by noting that, under the null hypothesis the star is 
constant, all orderings of the observations in time are statistically 
equivalent. The test is then
constructed by creating a large number of equivalent versions of the original
dataset by randomly shuffling the observations, and 
comparing the amplitude of the best-fitting sinusoid to the original data, with
the amplitudes of sinusoids fitted to shuffled data. If the true amplitude is
sufficiently remarkable compared to the artificial amplitudes (e.g. if the
true value is amongst the upper 0.1\% of artificial values), the star may be
considered a systematic variable. Such a permutation test is optimal in the 
sense that it is completely true to life under the null hypothesis: the 
permuted datasets have the same time points of observation, and the same
noise level, as the original data. There is but a single, but currently
insurmountable, problem: the excessive amounts of computer time required to
process the necessarily large number of replications of the original dataset.
We therefore do what we consider the next best thing, which is to use
the statistical properties of a large, representative sample for the Hipparcos
database of stars, to produce a selection criterion which ought to work
well in general.

The selection criterion is based on the signal-to-noise ratio $R$ of the
best-fitting sinusoid. The derivation of the exact form of the criterion is
empirical, rather than theoretical. Calibration of the criterion is based on 
the results for three large sample
datasets. In order to construct the latter, two non-overlapping sets of
10000 stars were randomly selected from the Hipparcos database. These lists
were supplemented by a third consisting of all the apparently constant 
Hipparcos stars brighter than a magnitude limit of $V=7$ ($\sim 10800$ stars). 
The epoch
photometry of each of the stars was extracted from the Hipparcos database. 
Suspect observations were removed in a two-step process: first, all 
measurements with Hipparcos flags larger than 7 were discarded. 
[Approximately 16\% of the {\it Hipparcos} photometric observations are
flagged, and 7.5\% have flag values larger than 2.  Generally speaking,
the higher the flag value, the less reliable the observation (ESA 1997,
Volume 3). Flag values less than 8 indicate that only one of the two {\it
Hipparcos} consortia accepted the particular observations. Flag values
greater than 7 indicate problems such as high background radiation, poor
pointing, contamination by other stars, etc.] If fewer than 
$N=20$ measurements remained, no analysis was attempted. Next, an 
iterative procedure was used to weed out outlying observations, by removing
all values further than 2.58$\sigma$ from the mean for that star. It is
entirely possible that the latter step discards viable measurements (e.g.
deep eclipses): however,
the intention was to retain only {\it typical} observations, i.e. to eliminate
observations which could exert undue influence on the results. Given the rather
modest sizes of the datasets for some stars, single atypical observations
can strongly affect results. 

A periodogram was calculated for each dataset over the frequency interval 
$[0,12]$ d$^{-1}$, as described in Koen (2001). The frequency corresponding
to the periodogram maximum was noted, and a sinusoid with this frequency fitted
to the data by linear least squares. The amplitude and phase of this sinusoid, 
together with the frequency, could then be used as starting guesses in a more
sophisticated nonlinear least squares determination of the three quantities.
The amplitude of the sinusoid and the standard deviation of the residuals were
noted; the ratio $R$ of the two is what we refer to as the ``signal-to-noise
ratio".

A plot of the signal-to-noise ratio $R$ against the number of
observations $N$ for all stars in each of the three samples shows an
apparent power law  dependence.  It is therefore natural to examine
the relationship between $\log R$ and $\log N$ (where "log" indicates
natural logarithms in this section of the paper only): the relevant
plots are in Fig. 1. Straight lines were fitted to each of the
datasets in Fig. 1, and the results are presented in Table 1.
Datasets containing small numbers of observations ($N<40$) were not
taken  into account, as the scatter for these is rather large;
furthermore, low  outlying observations in Fig. 1 ($\log R<-1.31$)
were discounted. The parameters of the three lines agree quite well.

The results in Table 1 are for the model
$$\log R=a+b\log N$$
or equivalently
$$R=e^a N^b \, .$$
The implication is that 
\begin{equation}
RN^{-b}={\rm constant}\, .
\end{equation}
where the exponent $b$ is in the range -0.468 to -0.460.
Non-parametric regression estimates of the mean values of $RN^{0.465}$, which 
should be very close to being constant if (1) holds, are
shown in Fig. 2. The method used to obtain the curves is known as ``loess", 
and consists in this instance of fitting locally linear estimates, with a
window width of included data of 60\% - see the brief discussion in 
Koen (1996), or the original papers by Cleveland \& Devlin (1988) and
Cleveland, Devlin \& Grosse (1988). In order to avoid distortions caused by
extreme points, data elements with  $|RN^{0.465}|>3$ were not taken into
account; this meant the exclusion of respectively 16, 13 and 12 points for
the three collections of datasets. The good qualitative agreement between
the curves for the three different collections of data shows that the 
exponent $b$ in (1) is not in fact perfectly constant. As will become clear,
the variations of the order of 0.22 in the mean value of $RN^{0.465}$, for
different $N$, could be of importance,
and need to be taken into account. We therefore work with
\begin{equation}
R_1=RN^{0.465}-M(N)
\end{equation}
where $M(N)$ is the mean of the three loess curves in Fig. 2.

Fig. 3 contains plots of the statistics $R_1$ for each of the three
collections of stars. It is unclear whether the larger scatter at smaller
$N$ is due to the larger number of datapoints, or whether the variance of
$R_1$ does in fact depend on $N$. Loess regressions of $R_1^2$ on $N$
were therefore carried out, omitting the same outlying points as in the
estimation of the mean. The results are given in Fig. 4. The agreement between
the three curves is gratifying, and implies a rapid rise in the variance
of $R_1$ as $N$ decreases below 100. The final form of the signal-to-noise
statistic is then
\begin{equation}
R_*=R_1/V^{1/2}(N)=V^{-1/2}(N)\left [ RN^{-0.465}-M(N) \right ]
\end{equation}
where $V(N)$ is the mean of the three loess curves in Fig. 4. The standardised
statistics $R_*$ are plotted in Fig. 5.

The percentiles of $R_*$ in Fig. 5 can now be used to produce critical values 
of this statistic. Of course, only the upper tail of the distribution is
of interest in this context. The percentage points are in Table 2, where the 
results for
each of the three individual collections of datasets are given for purposes
of comparison; the last column of the Table shows the percentage points 
derived from all the data -- these are the values used in what follows.

Inspection of the last column of Table 2 shows that relatively small
changes $\sim 0.3$ in $R_*$ are associated with relatively large
changes (factor $\sim 2$) in the significance levels of the statistic.
This underlines the necessity for the standardisation of $R$ into $R_*$.

One other very simple criterion is used to further weed out spurious 
variables. Many sets of observations have strong long-term trends, due to
slow aperiodic brightness variations, which could give rise
to prominent high frequency features in amplitude spectra through aliasing.
It is therefore also required that the identified period satisfies 
$$\Delta \equiv \frac{|P({\rm detrended})-P({\rm raw})|}
{\min [P({\rm detrended}), P({\rm raw})]} \le 0.001 = \Delta_c \, .$$
The detrending is performed as
described in Koen (2001); low order ($\le 3$) polynomials are fitted to the
data, and the fit with the highest significant order is used to prewhiten
the data. Results are virtually identical for critical values 
$\Delta_c$ in the range $10^{-5}$--0.005. 

The results of applying the criteria are displayed in Table 3,
which is based on selection with $R_* \ge 3.543$, i.e. the 0.1\% critical
value. There are
three groups of stars: those classified as constant, or unclassified, in
the Hipparcos catalogue; those classified as either ``unsolved" or
``microvariables"; and, for purposes of comparison, the Hipparcos ``periodic"
variables. It is instructive to consider results as a function
of magnitude for the former group, and this is done in the Table, and now
discussed. First, note
that the numbers of candidate variables selected by the $R_*$ criterion
exceeds the expected number of spurious selections by factors of the order
of 6.5--20 (although these numbers are misleading - see section 3). 
As expected, the percentage of variables decreases as the
magnitude limit rises. For this 
reason it was decided not to extend the search beyond $V>10$. Second, the
$\Delta$-criterion is obviously also quite stringent, particularly for
the brighter stars. As an example of the influence of the precise value
of $\Delta_c$, we note that changing it to $10^{-5}$ would have removed
one star from the final count in the top line in Table 3, while setting
$\Delta_c=0.005$ would have added four stars. 

For the sake of completeness we mention that of the 593 new variables
(i.e. stars in the first four lines of Table 3), 111 were classified as
constant (designation ``C") in the {\it Hipparcos} catalogue, while
484 were unclassified (variability field blank).

There is an encouraging agreement with the results in the Hipparcos catalogue,
in the sense that our $R_*$ criterion recovers 86\% of the Hipparcos periodic
variables. It is also noteworthy that the $\Delta$ criterion eliminates
only 16\% of the periodic variables, the corresponding number for the
unsolved variables being 56\%.

The properties of the 371 Hipparcos periodic variables rejected by the 
$R_*$ criterion were examined in some detail. Of these stars, 304 are
eclipsing binaries. For 96 stars periods were not determined from the
Hipparcos photometry [see the {\it Hipparcos} Variability Annex (ESA 1997, 
Volume 11)], while the periods
of 5 are shorter than our limit of 0.08 d. In total 313 (i.e. 84\%) of our
non-detections fell in at least one of these categories. Of the remaining 
light curves, the vast majority exhibit some or other aberration: sparse phase
coverage, low signal-to-noise, or unusual shapes (e.g. 
non-monotonic changes on the ascending or descending branches, or 
flat-bottomed minima with relatively sharp maxima).

The variance of the Hipparcos photometric measurement
errors was not constant with time (Eyer 1998), and thought
should be given to the possible implications for the method described above.
Since we have used an ordinary, rather than weighted, least squares algorithm, 
the only possible impact is through the initial frequency selection from the
periodogram. It is now shown that the frequency dependence of the first two
moments of the periodogram are unaffected by variability of the variance, and
hence that the choice of ``most likely" frequency is likewise unaffected.

Denote the deterministic (sinusoidal) signal by $f(t)$, and
the measurement errors by $e(t)$, such that E$e(t)\equiv 0$ and
var$[e(t)]=$E$e^2(t)=g(t)$, where $g(t)$ describes the time evolution of
the photometric error variance. It follows that
\begin{equation}
S(\omega)=\frac{1}{N}\left | \sum_t [f(t)+e(t)] \exp (-i\omega t) \right |^2
=S_f(\omega)+S_e(\omega)
\end{equation}
where $S(\omega)$, $S_f(\omega)$ and $S_e(\omega)$ are the periodograms
of the observations, the deterministic process, and the measurement errors
respectively. Now
\begin{eqnarray}
{\rm E}S_e(\omega)&=& \frac{1}{N}\left [\sum_t {\rm E}e(t) \cos \omega t 
\right ]^2 + \frac{1}{N}\left [\sum_t {\rm E}e(t) \sin \omega t \right ]^2\nonumber\\
&=&\frac{1}{N}\sum_t g(t)
\end{eqnarray}
and
\begin{eqnarray}
{\rm cov}[S_e(\omega),S_e(\psi)]&=&{\rm E}[S_e(\omega)S_e(\psi)]-
{\rm E}S_e(\omega) \, {\rm E}S_e(\psi)\nonumber\\
&=& \frac{1}{N^2} {\rm E}\sum_t e^2(t) \sum_j e^2(j)\nonumber\\
&-&\frac{1}{N^2} \left [\sum_t g(t) \right]^2 \nonumber\\
&=& \frac{1}{N^2} {\rm E}\sum_t e^4(t)-\frac{1}{N^2}
\left [\sum_t g(t) \right]^2 
\end{eqnarray}
provided that measurement errors at different epochs are uncorrelated.
If further $e(t)$ is independent of $f(t)$, the required result follows from
(4)--(6).

\section*{\large \bf 3 QUALITY CONTROL BY ADJUSTMENT OF SIGNIFICANCE LEVELS}
It is possible to adjust $R_*$ according to the brightness of the stars 
studied, in order to obtain homogeneously reliable results. The approach
is outlined below.

If all the stars were non-variable, the number selected by the $R_*$ criterion
would have had a binomial distribution: the probability of selecting $k$ stars
as variables, out of a sample of $N$ stars, is
\begin{equation}
{\rm Pr}(k)={N \choose k}p^k (1-p)^{N-k}\, ,
\end{equation}
where $p$ is the test level of $R_*$ (e.g. $p=0.001$ in Table 3). The numbers
in column 3 of Table 3 are the expected values $Np$ of spurious variables
for the $N$ given in
column 1 of the Table. The probability of selecting {\it at least} $K$
stars as variables is
$${\rm Pr}(k\ge K)=\sum_{k=K}^N {N \choose k}p^k (1-p)^{N-k}\, .$$
As an example, for the group of stars with $V \le 7$ ($N=10813$), the
probabilities of selecting at least 11, 21 or 38 stars as variables are
0.52, 0.002, and 10$^{-10}$ respectively. Clearly the probability of finding
as many as 244 candidate variables by chance is miniscule, or, conversely,
it is expected that most of the $R_*$-selected stars are truly variable.

The expected fraction $\eta$ of spurious variable stars amongst the $M$ 
candidate
variables selected by the $R_*$ criterion can be used as a measure of the
quality of the collection of candidates. Conditionally on the value of $M$,
$$\eta=\frac{Np}{M} \, .$$
The value of $\eta$ for the four brightness intervals in Table 3 cannot
be estimated from the numbers in the first four lines of the Table alone;
it is necessary to include candidate variables from the ranks of the
``unsolved" and ``periodic" Hipparcos stars. The results are in Table 4, for
three different significance levels of the $R_*$ criterion.

Clearly candidate variables selected on the basis of sufficiently large
$R_*$ have good probabilities of being true variables: for example, with 
Pr$(R_*)<0.001$, about 98 \% of the faintest group of candidates are expected
to be true variables. Nonetheless, the expected fraction of spurious variables
could differ by a factor of three for stars of different brightnesses. This
suggests adjusting $R_*$ to obtain homogeneous results. For example, in order
to have a uniform value of about 0.01 for $\eta$, $p=0.002$ could be used
for the brightest stars; $p=0.001$ for the group with $7 < V \le 8$; and
$p=0.0005$ for the stars fainter than $V=8$. 

A thorough implementation of such a ``quality control" scheme obviously
requires some further work, which is outside the scope of this paper.
Nonetheless, it should be clear that it is one of the potential advantages
of our variable selection methodology.

\section*{\large \bf 4 RESULTS}

The pertinent results for the newly selected candidate
periodic variable stars are presented in Table 5. 
Figs. 6a-e show an extract of the phase diagrams of
our candidates. The 5 pages are each composed of the phased data for the
first 36 stars in
the first 5 groups in Table 3.  Note that not all data
points from the Hipparcos epoch photometry are plotted, but only those 
selected as explained in section 2.

The periodic annex of the Hipparcos catalogue contains 2712 variables, selected
from a database of 118204 stars, giving a 2.3\% incidence of periodic
variables. In this study 2675 stars were selected from 94336 candidates.
Of course,
in order to compare this to the Hipparcos result, the information in the last
line of Table 3 should be incorporated, i.e. 4625 stars were selected from
97015 candidates, giving a percentage of 4.8.
As already pointed out the current aim is not the same, hence the difference
in results is not a cause for alarm.

Fig. 7 shows the period and amplitude distributions from Table 5, in the
form of an amplitude--period plot; for comparison, Fig. 8 contains the
corresponding results from the Hipparcos periodic variability annex.
The locations of the
Mira, Cepheid, RR Lyrae (ab and c types), and $\delta\,$Scuti stars are clearly
visible in the latter diagram.

In Fig. 7 we remark a strong accumulation of frequencies near
11.25 d$^{-1}$ (2h 08 min), which was the rotation frequency of the
satellite. These frequencies may be a cause for concern,
although the results are in the correct range of periods and amplitudes
for $\delta\,$Scuti stars. A quick look at the spectral types of the stars
confirms that many of the frequencies are probably spurious: for example,
many M giant stars have periods in the suspect range. As mentioned before,
there is a strong aliasing effect which is produced by convolution of low
frequencies with the spectral window. Although the $\Delta$-criterion of
section 2 was designed to remove such variables, it is evidently not
infallible. Furthermore, ``real" variability in the data due to the rotation
of the satellite remains a possibility: see Koen \& Schumann (1999) where
such an effect was shown to exist in Tycho epoch photometry.
In fact, the referee of this paper has pointed out that errors in the 
modelling of the background radiation (in the case of fainter stars) or
in the modelling of signal distortion (in the case of the brightest stars)
may give rise to spurious 11.25 d$^{-1}$ frequencies.

Table 6, which summarises the number distribution of 
variables in Table 5 as functions of frequency and spectral classification,
throws further light on the aliasing problem. First, the number distribution is
virtually constant for frequencies between 3 and 9.5 d$^{-1}$ (see the last
column in Table 6). The distribution increases sharply with higher frequencies,
reaching a peak in the bin $[11,11.5]$ d$^{-1}$. Second, in the frequency
range 6-10 d$^{-1}$, the majority of stars are of spectral types A and F,
and there are few late type stars. However, at higher frequencies, there is
a substantial excess of late type (particularly M) stars.

The high incidence of A and F type stars at high frequencies is to be expected:
these are the spectral types and frequencies associated with $\delta$ Scuti
stars, which are known to be very abundant. By contrast, the large number
of late type stars with high frequencies, is highly unexpected. There are
two obvious explanations: either there is a substantial aliasing problem for
these stars, or there is a class of rapidly variable late type stars which
has been overlooked in the past. Choosing between these alternatives is
beyond the scope of this paper.

 We note in passing that of the 1112 stars with periods in excess of 4 d,
only 43 have $P>500$ d.

The authors are only aware of one theoretical investigation of the
question of the correct determination of periods from {\it Hipparcos}
data, namely Eyer et al. (1994). Those authors studied the range
$0.03<P<1000$ d, and conclude that identifications are generally very
accurate once the signal to noise ratio exceeds 1.25. 
The results of recent simulation studies by ourselves support their
findings. On the other hand, van Leeuwen et al. (1997) claimed that
``...the sensitivity to detecting real periods in the range of a few days
to 100 days is very low". This is clearly a point deserving of further
study.

There is an aggregation of points around amplitudes of the order 
of 0.04 mag, and
periods of the order of 1.8 days, both in Fig. 7 and Fig. 8.
Classification of the stars in Table 5 into different variable types is beyond 
the scope of
this article, but looking at the spectral types, periods and
amplitudes, it is noticeable that many different phenomena could be at
work. Indeed,
stars of the following types could be present: $\alpha$ Can Ven, SX Ari, 
$\gamma\,$Dor, $\alpha$ Cyg, $\gamma$ Cas, BY Dra, FK Com, small amplitude
red variables, eclipsing binaries of all types, and slowly pulsating B stars. 
We note in passing that, as could have been anticipated, most amplitudes are 
small: only 6.5\% are larger than 0.1 mag. The smallest amplitude is 2.5 mmag.

In Fig. 9 finer detail such as an excess of periods 
near 57 days is visible. Now
56 days is the time interval during which the satellite
rotation axis described one revolution on the cone on which it precessed.
This seems to have generated an effect on the photometry of double stars,
which is confirmed by the
fact that the number of double star systems in the relevant histogram
bin is substantially greater than in the adjacent bins. 
In the Fig. 9 histogram bin containing the 58 day period, the fraction of 
double stars is 23\% (12 stars out of 53); by comparison the two lower adjacent 
bins, and the
two higher adjacent bins, have double star fractions 
of 7\% (4 stars out of 58) and 2\% (1 star out of 45) respectively.

\section*{\large \bf 5 CONCLUSIONS}
We conclude with some cautions:
It is important to bear in mind the precise property the $R_*$ statistic 
tests for, namely the existence of some frequency with which the data can be 
folded so that it shows an unusually large amplitude compared to the 
residual scatter. Although this will often mean that this frequency is truly
present in the data, it will not always be the case. Both scatter in the
measurements of constant stars, and fortuitous folding of the observations
of non-periodic stars, can lead to the spurious identification of periodic
variables, with data as sparse as those analysed here. Furthermore, the
sparsity, and particular time distribution of the observations, imply
that frequency aliasing is a substantial threat, so that all the identified
frequencies should be treated with caution. In particular, inspection of
Fig. 8 shows 
an excess of frequencies roughly in the range 10.5-11.5 d$^{-1}$, 
and it is well known (e.g. Eyer \& Grenon 2000)
that Hipparcos data are prone to aliasing of low frequencies to values near
11.2 d$^{-1}$.

On the other hand, for some Hipparcos datasets aliasing is, in practice,
minimal, and frequencies can be determined more easily than would have been the
case with typical groundbased observations - see, for example, the window
function of HD95321 shown in Koen et al. (1999). 

It must also be borne in mind that there are classes of periodic variables
(e.g. $\delta$ Scuti stars) which commonly have frequencies beyond our
detection limit of 12 d$^{-1}$. We will either have failed to identify
such stars as variables, or will have found aliases of the true frequencies.

Subject to all the above qualifications, we note that the value of the $R_*$
statistic can be used to classify
the candidates in order of ``significance". In other words, this statistic
renders possible a comparison between stars of different magnitudes
and of different numbers of measurements.

\begin{table*}
\caption{The results of fitting straight lines to each of
the three datasets in Fig. 1. Only signal-to-noise ratios in excess of 0.27
($\log R>-1.31$) and datasets containing at least 40 observations 
($\log N>3.69$) were taken into account in the fitting. Standard errors
of the estimates are given in brackets.}

\begin{tabular}{cccc}
Dataset  & Intercept & Slope  &  $N$  \\
    & & & \\
 1 & 1.71 (0.016) & -0.466 (0.0035) & 9684 \\
 2 & 1.68 (0.016) & -0.460 (0.0036) & 9732 \\
 3 & 1.72 (0.016) & -0.468 (0.0034) & 10746\\[2cm]
\end{tabular}
\end{table*}

\begin{table*}
\caption{Percentage points of the statistic $R_*$, for each of
the three collections of stars, and for the three datasets combined.}

\begin{tabular}{cccccc}
Dataset  & 1 & 2 & 3 & & All  \\
    & & & & & \\
 N     & 9747 & 9789 & 10813 & & 30349 \\
    & & & & & \\
 1\%   & 2.56 & 2.59 & 2.55  & & 2.566\\
0.5\%  & 2.92 & 2.90 & 2.83  & & 2.888\\
0.2\%  & 3.44 & 3.24 & 3.24  & & 3.282\\
0.1\%  & 3.58 & 3.42 & 3.51  & & 3.543\\
\end{tabular}
\end{table*}

\pagebreak

\begin{table*}
\caption{Application of the selection criteria to stars classified
as constant (or unclassified), and those classified as ``unsolved" (or 
``microvariable"). The former group of stars has been subdivided
according to brightness. The column headed ``Pr$(R_*)<0.001$" shows
the number of stars with $R_*>3.543$ (the 0.1\% point) from each grouping;
the column headed ``Final" are the numbers finally accepted as
variables. Results for the Hipparcos ``periodic" variables are also shown,
for purposes of comparison.}

\begin{tabular}{cccccc}
Dataset & No. candidates & Expected spurious & Pr$(R_*)<0.001$ & 
 $\Delta>0.001$ & Final \\
    & & & &  &\\
$V \le 7$   &10813 & 11 & 244 & 95 & 149\\
$7<V\le 8$  &20149 & 20 & 224 & 77 & 147\\
$8<V \le 9$ &34411 & 34 & 327 &114 & 213\\
$9<V\le 10$ &20134 & 20 & 154 & 70 & 84\\
 Unsolved & 7784  &    &4396  &2493& 1908\\
Micro& 1045 &  & 313 & 139 & 174\\
  & & & & & \\
 TOTALS: & 94336 &  & 5658 & 2988 & 2675 \\ 
  & & & & & \\
 Periodic & 2679  &    &2308 &   359& 1949\\[2cm]
\end{tabular}
\end{table*}

\pagebreak
\begin{table*}
\caption{First page of the 2675 periodic candidate variable list.}
\begin{tabular}{rrrrrrrrrrl}
\hline
HIP&V&FREQ&AMPL&N2&R&SpTy&H53\\
\hline
38& 8.65&0.70787&0.0123&125&4.158&G6V &C\\
102& 7.05&0.07586&0.0069&195&6.522&M1III & \\
274& 6.24&0.34712&0.0166&130&6.261&B3Ia &U\\
279& 8.69&0.61426&0.0155&128&5.990&B8V & \\
281& 8.45&11.08512&0.0148&152&3.811&M0 &U\\
283& 7.45&6.16640&0.0133&88&5.313&A5 &U\\
292& 8.11&0.37690&0.0122&113&5.783&A0 &U\\
355& 4.99&11.22171&0.0053&132&4.968&K3Ibvar &M\\
457 & 7.47&0.00199&0.1044&172&19.893&M3III &U\\
458& 6.97&11.44378&0.0081&99&4.911&K5 & \\
519& 7.92&1.47910&0.0740&119&7.687&F3IV &U\\
536& 9.76&0.00361&0.2017&89&15.533&M6e &U\\
605& 7.48&0.80210&0.0136&149&4.639&K5III &U\\
621& 7.50&0.10826&0.0249&142&11.723&M1III &U\\
632& 9.06&0.06260&0.0351&165&12.781&M2III: &U\\
690& 9.26&0.05131&0.0143&134&3.736&G3/G5IV/V & \\
720& 7.13&0.25630&0.0277&102&7.893&M2III &U\\
852& 7.11&0.01851&0.0410&97&7.574&M4III: &U\\
893& 7.78&0.03264&0.1070&101&10.773&M3III &U\\
926& 7.47&0.08279&0.0080&130&5.620&K0 & \\
949& 8.09&0.55897&0.0094&195&4.308&B9 & \\
967& 6.18&0.73010&0.0058&153&4.786&K4III &M\\
970& 7.80&11.65544&0.0202&160&7.867&M1III &U\\
989& 7.69&0.31618&0.0732&126&14.190&M... &U\\
999& 8.44&0.52745&0.0227&121&6.264&K0 &U\\
1024& 9.55&1.88707&0.0151&115&4.475&A5 & \\
1035& 8.84&0.52293&0.0100&149&3.657&F5 & \\
1086& 5.71&0.85784&0.0062&173&10.890&F0IV &M\\
1124& 7.17&0.13408&0.0157&110&4.216&M4III: &U\\
1131& 8.25&0.48166&0.0380&129&7.146&M2 &U\\
1146& 6.63&0.10180&0.0496&145&15.551&M1III &U\\
1158& 5.13&0.45740&0.0213&66&7.372&M3IIIvar &U\\
1168& 4.79&0.16767&0.0094&104&6.427&M2III &U\\
1191& 5.77&2.49699&0.0077&75&3.958&B9V &M\\
1289& 8.83&0.05126&0.0182&181&4.759&M0 &U\\
1551& 7.45&0.20012&0.0165&116&4.190&M3III &U\\
1555& 7.78&0.04424&0.0422&153&21.399&M1III &U\\
1571& 7.44&0.04682&0.0257&229&21.949&M1/M2III &U\\
1609& 7.53&0.10716&0.0364&101&13.501&M0 &U\\
1623& 7.87&0.71752&0.0117&115&4.240&F6V & \\
1629& 7.66&1.31813&0.0121&146&5.470&F2 &U\\
1652& 6.80&0.89264&0.0212&121&6.295&M2II: &U\\
1655& 8.38&0.08694&0.0305&101&8.871&G5III &U\\
1763& 8.76&0.26470&0.0256&61&4.051&M0 &U\\
1792& 7.94&0.17663&0.0215&73&5.145&G5 &U\\
1843& 9.99&0.78347&0.0251&97&3.682&M0 &U\\
1880& 7.70&11.72517&0.0267&155&8.822&M2III &U\\
1941& 7.55&0.24823&0.0323&82&6.593&M... &U\\
1945& 9.26&11.58187&0.0291&208&6.453&C4.5v &U\\
2086& 6.24&0.08465&0.0153&86&3.589&M1III &U\\
2164& 8.01&0.23288&0.0225&97&7.334&M2/M3III &U\\
2203& 8.64&3.75862&0.0147&81&5.648&K2III &C\\
2219& 5.01&0.21860&0.0830&67&14.094&M3IIIvar &U\\
2225& 5.18&0.45460&0.0044&154&7.629&A2Vs & \\
2254& 9.98&1.17703&0.0118&96&3.785&G2 & \\
2283& 7.35&0.58540&0.0182&91&6.738&K5III &U\\
2285& 9.01&0.14403&0.0351&143&13.439&M0 &U\\
2340& 7.93&0.24137&0.0345&87&6.372&K2 &U\\
2388& 6.18&7.37913&0.0104&156&7.612&F2III &U\\
2474& 6.18&0.92717&0.0068&137&4.190&B6V &M\\
\hline
\end{tabular}
\end{table*}

\pagebreak
\begin{table*}
\caption{A check on the reliability of the $R_*$ selection criterion:
the parameter $\eta$ is the expected percentage of false variables amongst
the $M$ selected stars. The numbers in column 2 refer to {\it all} Hipparcos 
stars in the particular magnitude interval.}

\begin{tabular}{cccccccccc}
        &                &\multicolumn{2}{c}{$p=0.002$}&
  &\multicolumn{2}{c}{$p=0.001$}&
  &\multicolumn{2}{c}{$p=0.0005$}\\
Dataset & No. candidates &  $M$ & $\eta$ && $M$ & $\eta$ && $M$ & $\eta$ \\
    & & & & & & && &\\
$V \le 7$   &13715 & 2371 & 1.16 && 2233 & 0.61 && 2119 & 0.32\\
$7<V\le 8$  &22875 & 2129 & 2.15 && 1984 & 1.15 && 1860 & 0.61 \\
$8<V \le 9$ &37399 & 2208 & 3.39 && 2040 & 1.83 && 1888 & 0.99\\
$9<V\le 10$ &21777 & 1149 & 3.79 && 1071 & 2.03 && 988  & 1.10 

\end{tabular}
\end{table*}

\pagebreak

\begin{table*}
\caption{The number distribution of the candidate variables in Table
5, as a function of frequency and spectral type. 
Classifications  R, N, S and C are included in under M, while 
``Other'' comprises primarily unclassified stars and composite spectra.}

\begin{tabular}{ccccccccccc}

& \multicolumn{8}{c}{Spectral Type} & & \\
Frequency Interval  & O & B& A & F & G &  K & M  &   Other & & Total\\
 & & & & & & & & & \\
$[11.5,12.0]$& 1 &   4&    9 &   8   &   7  &    18  &  22  &         & &  69\\
$[11.0,11.5]$& 2 &  15&   12 &   7   &   6  &    39  &  63  &     5   & & 149\\
$[10.5,11.0]$& 0 &   8&    6 &   8   &   5  &    18  &  17  &     1   & &  63\\
$[10.0,10.5]$& 0 &   6&    3 &   4   &   5  &     5  &   3  &         & &  26\\
 & & & & & & & & & &\\
$[ 9.5,10.0]$& 0 &   4&    7 &   8   &   0  &     4  &   0  &         & &  23\\
$[ 9.0, 9.5]$& 0 &   4&    6 &   4   &   0  &     1  &   0  &         & &  15\\
$[ 8.5, 9.0]$& 0 &   2&    6 &  10   &   1  &     0  &   0  &         & &  19\\
$[ 8.0, 8.5]$& 0 &   0&    2 &   6   &   1  &     0  &   0  &         & &   9\\
 & & & & & & & & & &\\
$[ 7.5, 8.0]$& 0 &   2&    2 &   6   &   2  &     3  &   0  &         & &  15\\
$[ 7.0, 7.5]$& 0 &   1&    1 &   7   &   2  &     0  &   0  &         & &  11\\
$[ 6.5, 7.0]$& 0 &   0&    5 &   7   &   0  &     1  &   0  &         & &  13\\
$[ 6.0, 6.5]$& 0 &   1&    5 &   9   &   0  &     0  &   0  &         & &  15\\
 & & & & & & & & & &\\
$[ 5.5, 6.0]$& 0 &   2&    3 &   4   &   3  &     2  &   0  &         & &  14\\
$[ 5.0, 5.5]$& 0 &   3&    1 &   5   &   3  &     4  &   0  &         & &  16\\
$[ 4.5, 5.0]$& 0 &   2&    6 &   3   &   2  &     0  &   0  &         & &  13\\
$[ 4.0, 4.5]$& 0 &   4&    3 &   3   &   0  &     0  &   0  &         & &  10\\
 & & & & & & & & & &\\
$[ 3.5, 4.0]$& 0 &   3&    4 &   1   &   1  &     3  &   0  &         & &  12\\
$[ 3.0, 3.5]$& 1 &   7&    3 &   4   &   1  &     1  &   0  &         & &  17\\
$[ 2.5, 3.0]$& 0 &   2&    9 &   5   &   1  &     5  &   0  &         & &  22\\
$[ 2.0, 2.5]$& 0 &  11&    9 &   6   &   4  &     6  &   0  &     1   & &  37\\
 & & & & & & & & & &\\
$[ 1.5, 2.0]$& 0 &  18&   24 &  19   &   7  &     5  &   4  &         & &  77\\
$[ 1.0, 1.5]$& 0 &  57&   34 &  21   &   6  &    13  &  30  &     4   & & 165\\
$[0.75, 1.0]$& 0 &  40&   21 &  18   &   4  &    19  &  30  &     2   & & 134\\
$[0.50,0.75]$& 0 &  57&   33 &  19   &   9  &    35  &  72  &     4   & & 229\\
 & & & & & & & & & &\\
$[0.25,0.50]$& 3 &  81&   32 &  14   &  17  &    68  & 162  &    13   & & 390\\
$[0.00,0.25]$& 5 & 117&   59 &  35   &  73  &   243  & 550  &    30   & &1112

\end{tabular}
\end{table*}

\pagebreak
\begin{figure*}
\resizebox{\hsize}{!}{\includegraphics{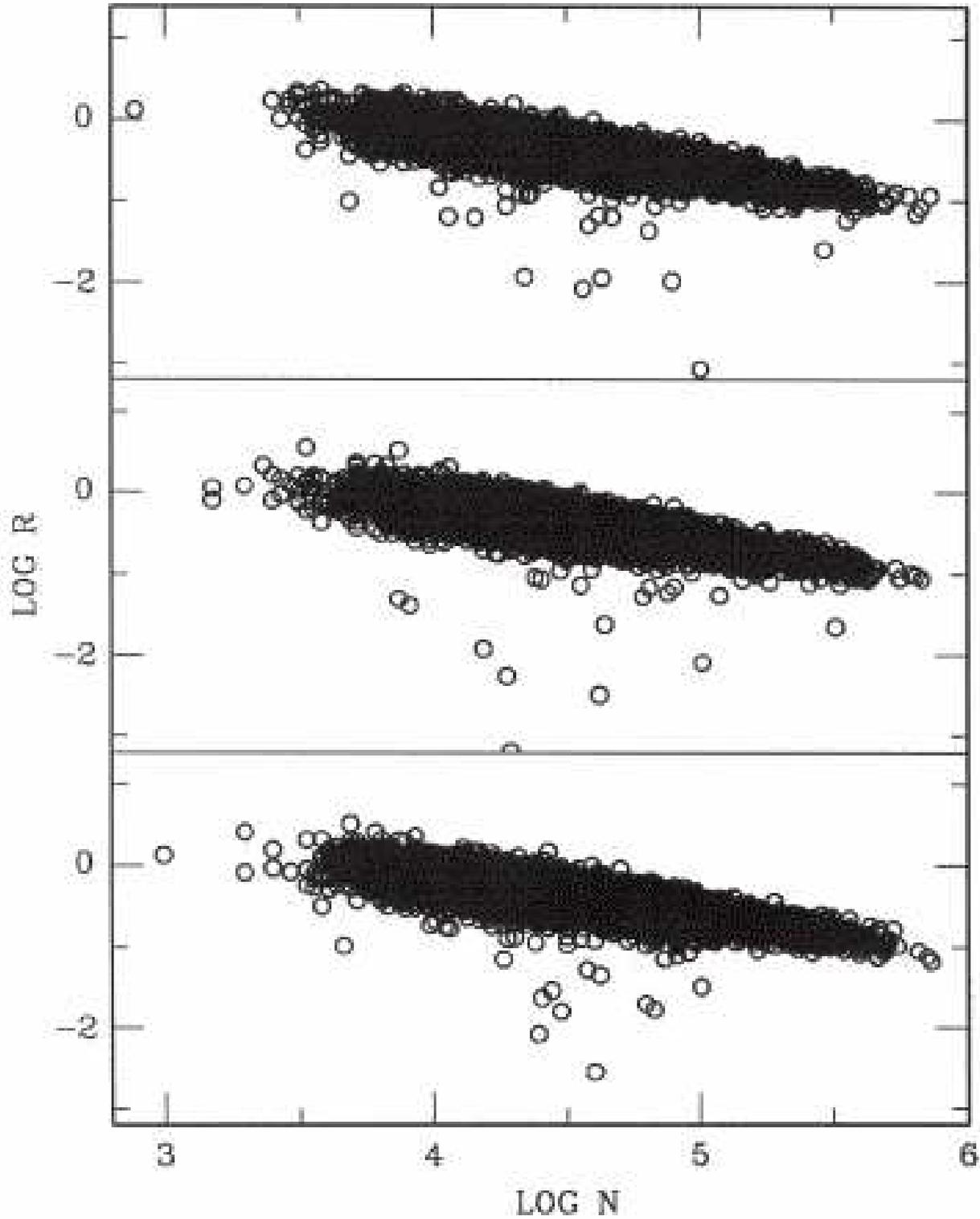}}
\caption{A log-log plot of the signal-to-noise ratio $R$ against the number
of accepted observations of each star, for each of the three collections 
of roughly 10000 stars.}
\end{figure*}

\begin{figure*}
\resizebox{\hsize}{!}{\includegraphics{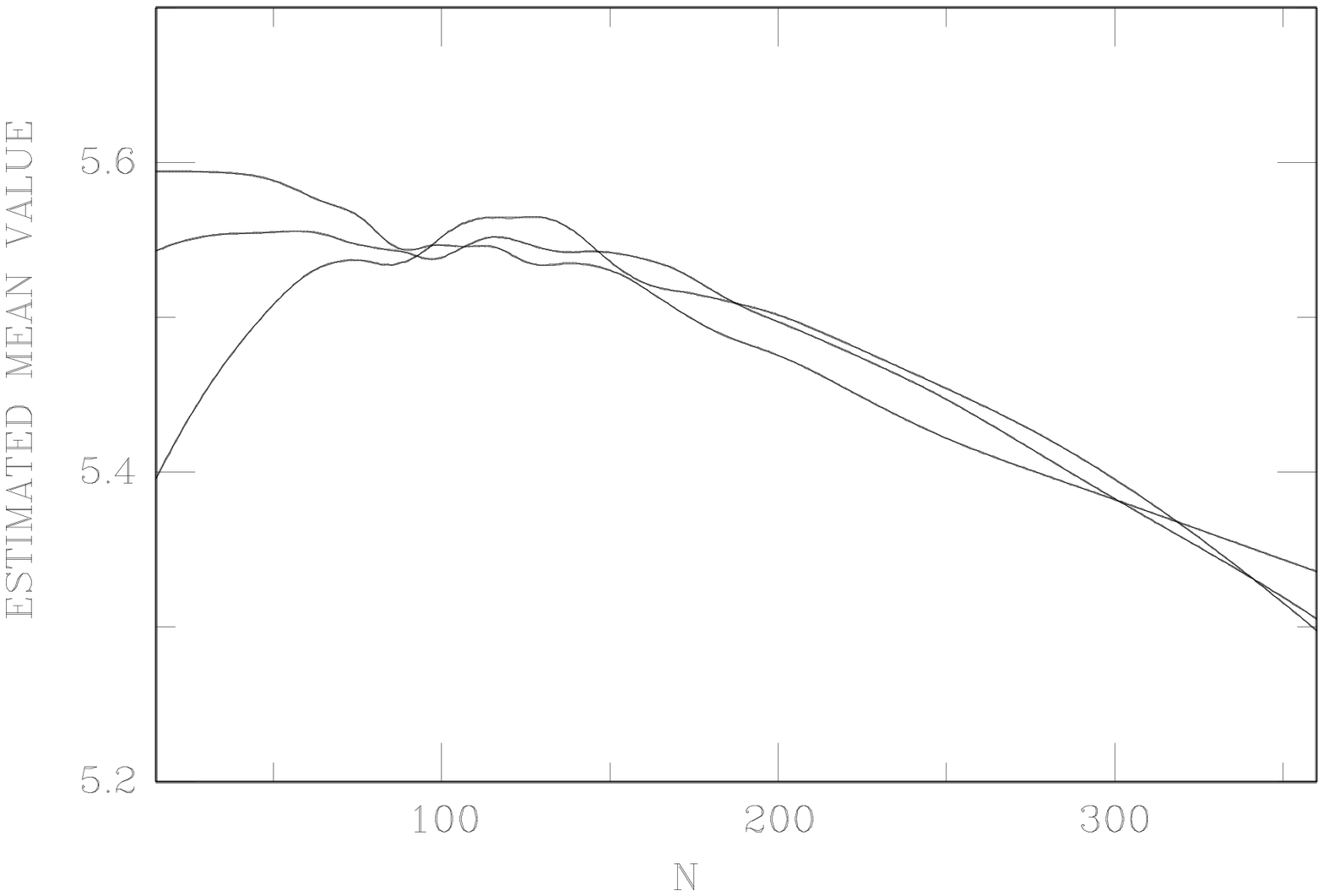}}
\caption{Non-parametric regression estimates of the means of $RN^{0.465}$, for
each of the three collections of stars.}
\end{figure*}

\begin{figure*}
\resizebox{\hsize}{!}{\includegraphics{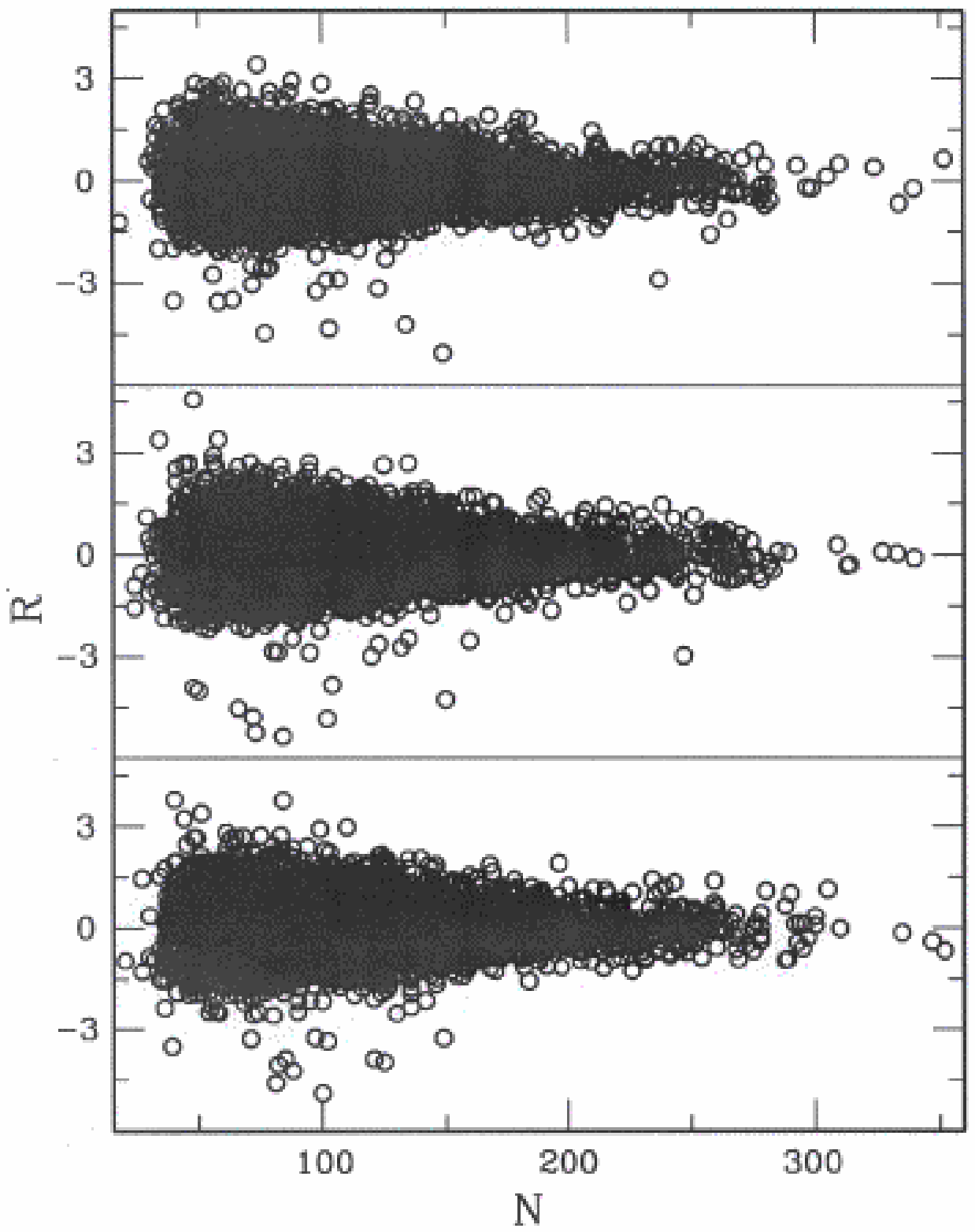}}
\caption{The statistic $R_1$ (see Eqn. (2)) for all stars in each of
the three collections.}
\end{figure*}

\begin{figure*}
\resizebox{\hsize}{!}{\includegraphics{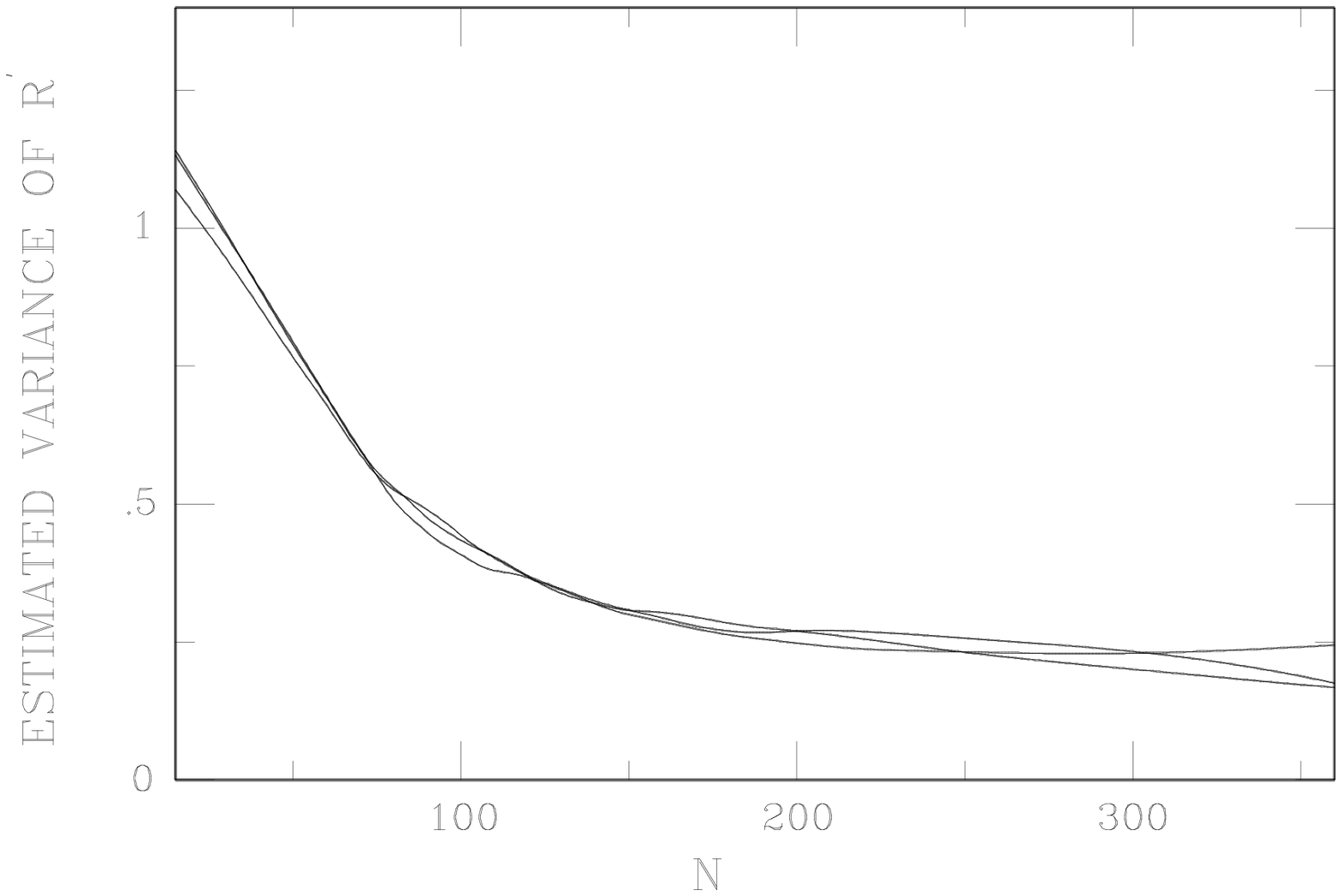}}
\caption{Non-parametric regression estimates of the variances of $RN^{0.465}$, 
for each of the three collections of stars.}
\end{figure*}

\begin{figure*}
\resizebox{\hsize}{!}{\includegraphics{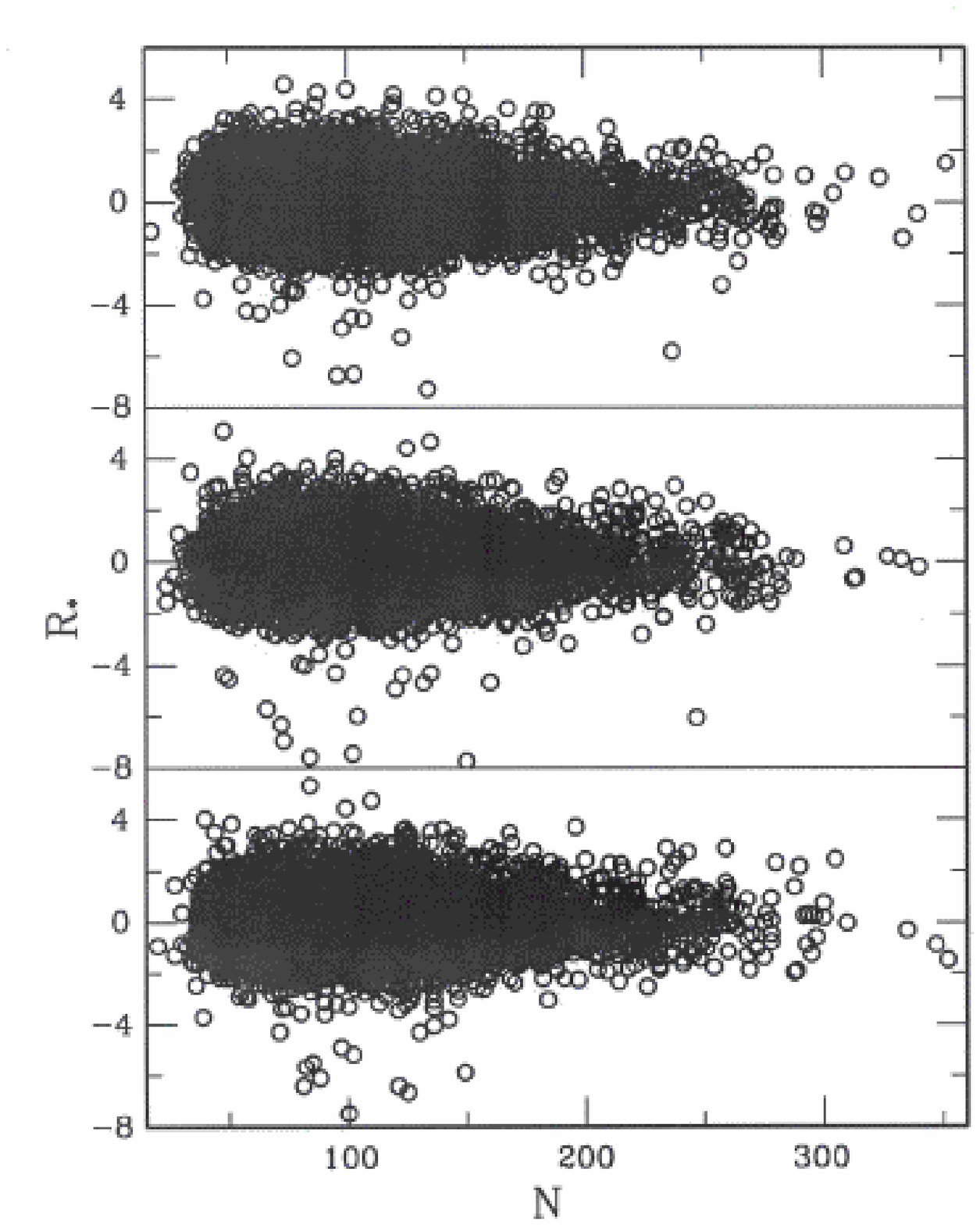}}
\caption{The statistic $R_*$ (see Eqn. (3)) for all stars in each of
the three collections.}
\end{figure*}

\begin{figure*}
{\includegraphics[height=0.9\textheight]{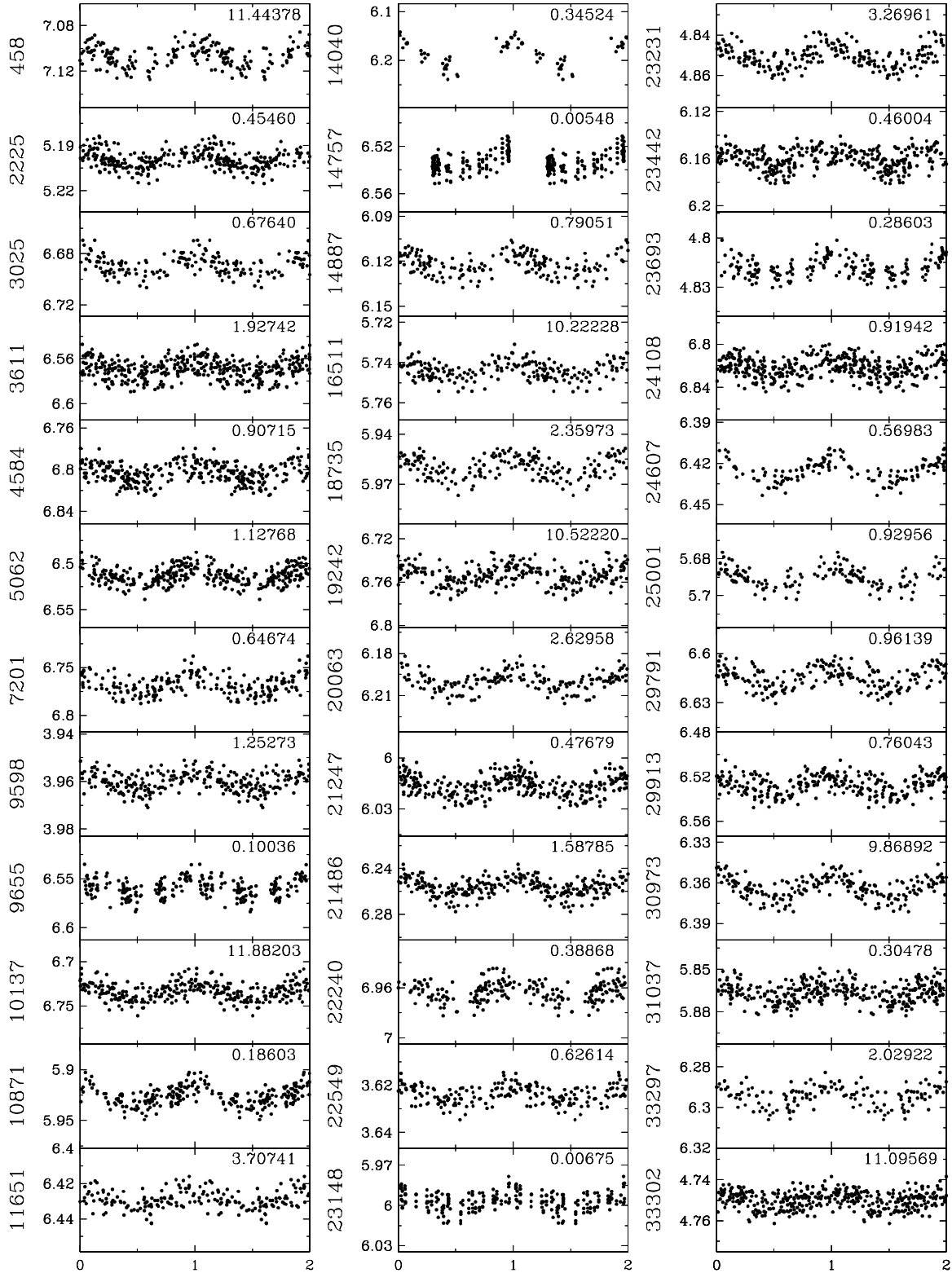}}

\caption{[Note: Only Fig. 6a is displayed here] Phased lightcurves for a sample of the candidate periodic variables 
in Table 5, for stars with $V\le 7$ (a); $7 < V \le 8$ (b); $8< V \le 9$ (c);
$9 < V \le 10$ (d); and for stars classified as ``unsolved" variables
in the Hipparcos catalogue (e). The customary two cycles of variation are 
plotted.}
\end{figure*}

\begin{figure*}
\resizebox{\hsize}{!}{\includegraphics{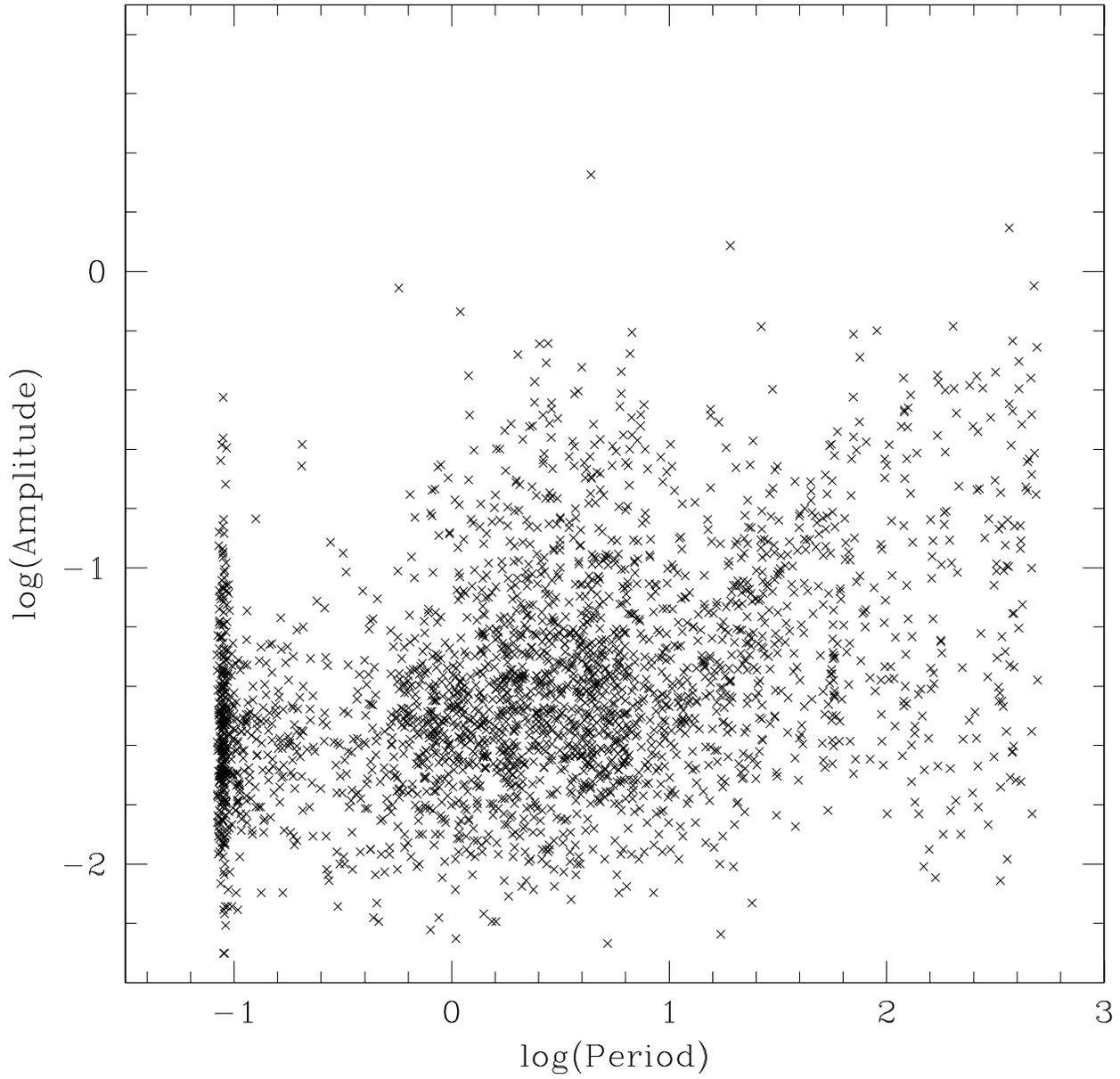}}
\caption{Amplitudes and frequencies of the new candidate variables in Table
5.}
\end{figure*}

\begin{figure*}
\resizebox{\hsize}{!}{\includegraphics{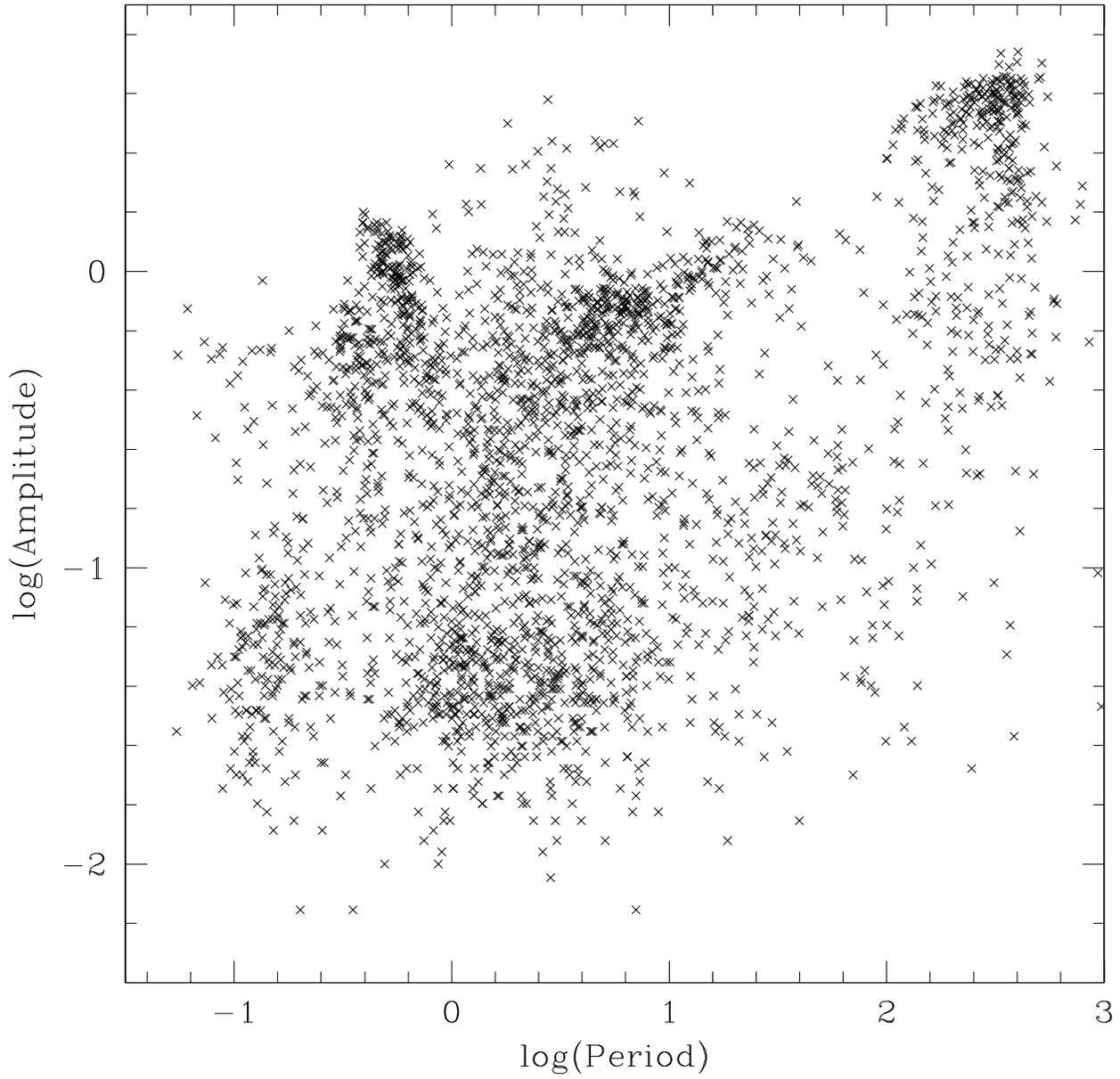}}
\caption{Amplitudes and frequencies of the Hipparcos periodic variables.}
\end{figure*}

\begin{figure*}
\resizebox{\hsize}{!}{\includegraphics{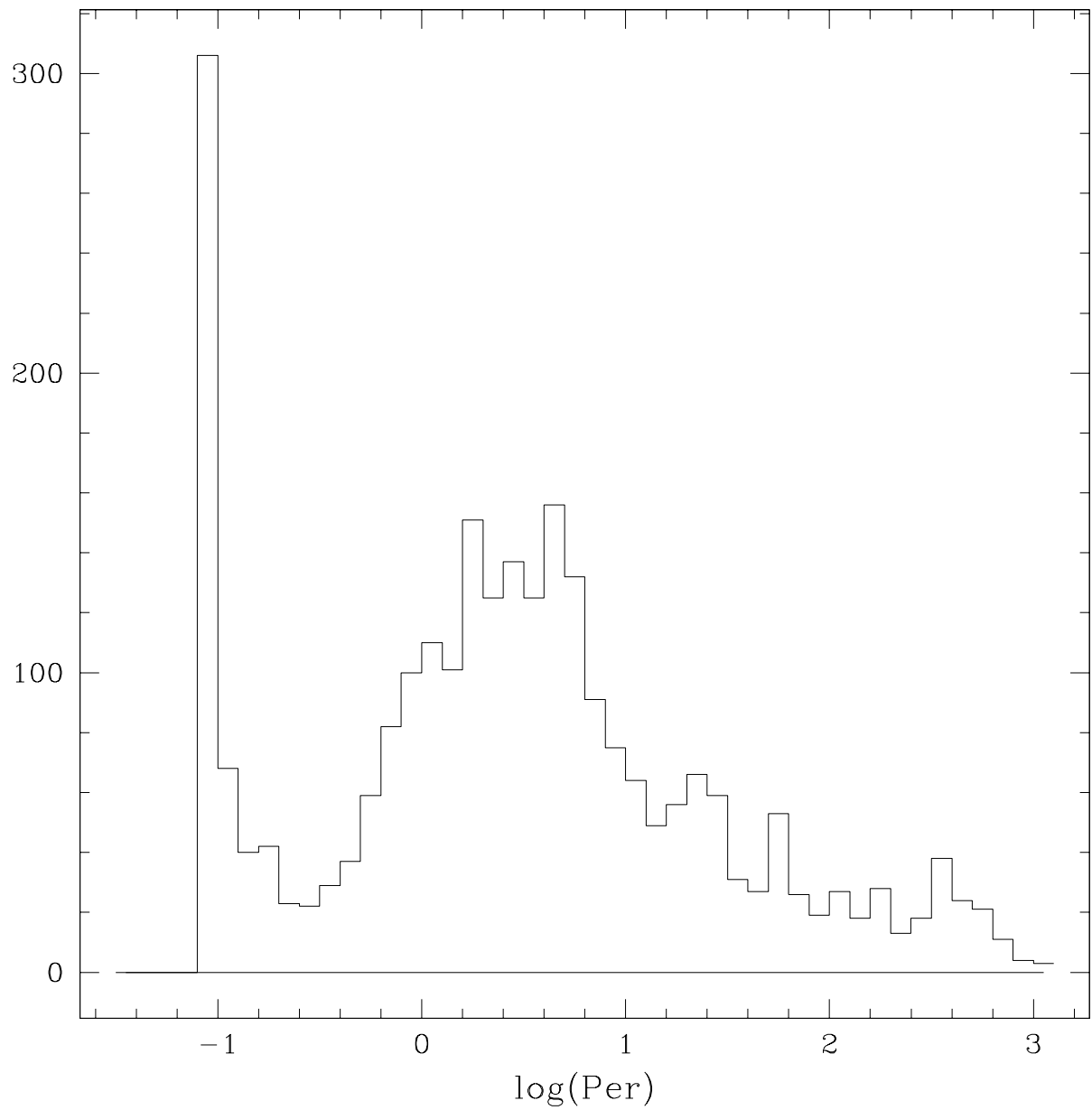}}
\caption{The distribution of the frequencies of the new candidate
periodic variables in Table 5.}
\end{figure*}

\end{document}